Milan M. Ćirković
Astronomical Observatory
Belgrade, Serbia

Slobodan Perović
Department of Philosophy
University of Belgrade
Serbia

# ALTERNATIVE EXPLANATIONS OF THE COSMIC MICROWAVE BACKGROUND: A HISTORICAL AND AN EPISTEMOLOGICAL PERSPECTIVE

**Abstract:** We historically trace various non-conventional explanations for the origin of the cosmic microwave background and discuss their merit, while analyzing the dynamics of their rejection, as well as the relevant physical and methodological reasons for it. It turns out that there have been many such unorthodox interpretations; not only those developed in the context of theories rejecting the relativistic ("Big Bang") paradigm entirely (e.g., by Alfvén, Hoyle and Narlikar) but also those coming from the camp of original thinkers firmly entrenched in the relativistic milieu (e.g., by Rees, Ellis, Rowan-Robinson, Layzer and Hively). In fact, the orthodox interpretation has only incrementally won out against the alternatives over the course of the three decades of its multi-stage development. While on the whole, none of the alternatives to the hot Big Bang scenario is persuasive today, we discuss the epistemic ramifications of establishing orthodoxy and eliminating alternatives in science, an issue recently discussed by philosophers and historians of science for other areas of physics. Finally, we single out some plausible and possibly fruitful ideas offered by the alternatives.

## 1. Introduction

The discovery of the cosmic microwave background (CMB) in 1965 by Arno Penzias and Robert Wilson and interpreted by Robert H. Dicke and his co-workers was a turning point in 20th century cosmology. It divided cosmology into an epoch of sometimes heated *cosmological controversy* (Kragh 1996) and an epoch of solidified support for the standard cosmological paradigm, popularly known as the *hot Big Bang cosmology* (Peebles, Page, and Partridge 2009). Actually, attributing the discovery of the CMB to Penzias and Wilson is a bit misleading, first, because they were not looking for it and, second, because it had been predicted by Gamow and his collaborators a few decades earlier.[1] They initially interpreted the accidentally detected signal as a noise caused by an artefact; they were not aware it had anything to do with a physical phenomenon of the utmost importance for cosmology. Their detection of the signal had far-reaching implications, however, not least of which was a now overlooked interpretation race in which they themselves did not participate.

The fact that the 1965 discovery was a clear watershed creates the impression of inevitability of the currently standard interpretation of the great CMB discovery as a remnant of primordial fireball, and that no alternative interpretations have been offered, seriously or half-seriously, by distinguished cosmologists. The impression of the inevitability of the current view is shared by astronomers and laypersons alike. Two of the best cosmology textbooks available, by Coles and Lucchin (1995) and Peacock (1999), reinforce this impression. Peacock even notes, with a poetic flourish, "The fact that the properties of the last-scattering surface are almost independent of all the unknowns in cosmology is immensely satisfying, and gives us at least one relatively solid piece of ground to act as a base in exploring the trackless swamp of cosmology" (p. 290).

From the point of view of the astrophysics community, the validity of the orthodox interpretation of CMB is largely resolved, with some doubts voiced from time to time (e.g., Baryshev, Raikov and Tron 1996). And as far as the general issue of the choice of cosmological models is concerned, the standard cosmological model seems to rest on a secure foundation (for review of some exotic alternatives, see Ellis 1984).

Yet López-Corredoira (2014) has quite recently examined some alternative cosmological models from a sociological point of view. This is important, as the emergence of alternatives and their destiny is a complex issue at the heart of scientific knowledge production and the discovery process. For instance, Cushing (1994) argues that a perfectly viable alternative to the Copenhagen interpretation of Quantum Mechanics, Bohm's mechanics, has been side-lined because it was devised later on. And Chang (2010, 2009) says forgotten and abandoned alternatives are often alternate routes to discoveries that were never taken. He demonstrates this using relevant examples in chemistry. Perovic (2011) analyses how subtle changes in experimental conditions influence the possibility of emerging and often crucial alternative theoretical accounts in particle physics, while Dawid, Hartman, and Sprenger (2014) offer a Bayesian analysis of theoretical preferences when viable theoretical alternatives are not available.

---

[1] There are claims of earlier CMB detections, as described Peebles et al. (2009). Normative understanding of scientific discovery correctly rejects such claims in the same manner as we reject the idea that Galileo discovered Neptune, although he did observe it in 1612-13, giving the credit to Le Verrier in 1846.

The CMB is another case, and in many respects, a different and fruitful case, the study of which can enrich this strand of methodological and philosophical research. Generally speaking, in the scientific fields that reconstruct evidence from observations, the epistemic standing of orthodox thought is tied to the epistemic standing of available alternatives. Evidence in such cases is, on the whole, very different from evidence provided in, say, experiments in solid state physics, in the sense that the underdetermination of theoretical accounts by evidence is bound to be much more pronounced and longer lasting. The wiggle room for alternative interpretations is much wider in a field such as cosmology than in experimental physics, as the latter provides much more direct evidence in debates and thus severely constrains theoretical accounts of relevant phenomena. The CMB was a milestone discovery, but it would be misleading to think it played a role identical to that, for instance, played by the evidence delivered by a particle collider in competing theoretical approaches to the existence of an elementary particle . Its role unravelled much more gradually.

Given this, it is wise to avoid treating side-lined alternative interpretations in the same way as we justifiably would  experimentally falsified alternatives in experimental physics. Instead, we should generally regard them as a resource that can potentially be revised and revived (despite occasional fairly straightforward falsifications of its certain aspects) The evidence of orthodoxy does not necessarily justify our outright discarding of the alternatives in cosmology. In fact, establishing orthodoxy may unjustifiably boost the CMB's epistemic standing by eliciting ignorance or a too-hasty dismissal of the existing alternatives, in part by propagating an inadequate history of the field and systematically, albeit unjustifiably, downplaying existing alternatives. Failing to understand the subtleties of the history of how orthodox thought about CMB was established runs the risk of generating widespread prejudice that opinions dissenting from the standard paradigm are both few and insignificant.

In short, the CMB provides an incentive for philosophically minded historical research. Just how convincing was the account that became the standard CMB interpretation in the first years after Penzias' and Wilson's discovery or during the first decade or two thereafter? Were any viable alternatives neglected at the time? How convincing is the account currently, and are there any viable alternatives now? Has there been enough critical examination in the modern practical work on the issue? All these questions are part of the complex and insufficiently studied problematic of paradigm formation in modern cosmology (Kragh 1997). In the first part of the paper (Sections 2, 3, and 4), we offer a historical case study of the formation of the alternatives in modern cosmology, setting the basis for an assessment of their respective epistemic standing in the second part of paper (Sections 5 and 6).

Peebles' (1999) commentary on the centennial re-edition of Penzias and Wilson (1965) paper is a good starting point for our research into the historical context of the CMB:

> A willingness to believe such an elegant gift from nature surely also played a significant role in the early acceptance of the CBR [cosmic background radiation] interpretation... During four decades of involvement with this subject, I have grown used to hearing that such advances have at last made cosmology an active physical science. I tend to react badly because I think cosmology has been an active physical science since 1930, when people had assembled a set of measurements, a viable theoretical interpretation, and a collection of open issues that drove further research. This equally well describes cosmology today.

This comment sets the stage for the article. The "willingness to believe" the standard model and a lack of confidence in the seriousness of the pre-1965 cosmological research are key ingredients in the standard, streamlined view of the history of physical cosmology. There is a widespread impression that the microwave noise detected serendipitously by Penzias and Wilson threw us into an epoch of serious, quantitative cosmology and that the essential validity of the hot Big Bang paradigm has remained unchallenged ever since. As Coles and Lucchin (1995) suggest, "it is reasonable to regard this discovery as marking the beginning of 'Physical Cosmology'" (p. xiii).

Yet the impression is wrong and creates a false picture of both the history and the methodology of cosmology. The facts about multiple methodologically sound alternative explanatory hypotheses of the CMB are mostly forgotten. Consequently, important historic-philosophical lessons about contemporary cosmological research are missed, and a source of potentially valuable ideas side-lined. It is worth trying to weave a historical tapestry of this admittedly amazing development by considering some strands presently deemed peripheral. The general motivation for this study is perhaps best expressed by Helge Kragh's (1997) comments on the history of cosmology:

> There is a tendency to streamline history and ignore the many false trails and blind alleys that may seem so irrelevant to the road that led to modern knowledge. It goes without saying that such streamlining is bad history and that its main function is to celebrate modern science rather than obtain an understanding of how science has really developed. The road to modern cosmology abounded with what can now be seen were false trails and blind alleys, but at the time were considered to be significant contributions.

The story of the CMB alternative interpretations is paradigmatic in this respect. Many scientists and popularizers of science, perhaps justifiably, use every opportunity to hail the orthodox interpretation of CMB as one of the greatest, often as *the* greatest triumph of modern cosmological science. Yet its often professed role in terminating the cosmological controversy blurs the distinction between the physical phenomenon and the historical role of the dominant interpretation, ascribing some form of "progressive" value to the CMB photons themselves. The necessary palliative is certainly the study of the non-standard, minority interpretations which challenged the prevailing orthodoxy. In addition, as frequently happens in such circumstances, alternative theories may contain valuable side ideas, motivations, and conjectures. Because these theories are usually regarded as failures, their insights are understandably overlooked. This is actually quite common in the history of physics. For instance, in the cases of Machian theories of gravitation, such as Brans-Dicke theory (e.g., Dicke 1962) or Wheeler-Feynman action-at-a-distance classical electrodynamics (Wheeler and Feynman 1945, 1949; Hoyle and Narlikar 1964, 1971; Hogarth 1962), we encounter concepts too radical for their epoch, but which have since become the focus of debates in inflationary cosmology or of philosophical discussions on the arrow of time (Linde 1990;

Price 1991).[2] Such cases offer an additional pragmatic argument for studying well-motivated unorthodoxies in their own right.

Finally, from a broader point of view of criticisms of cosmology in general, a constant feature of 20[th] century science (e.g., Dingle 1954; Disney 2000), the issue of the epistemological significance of the CMB is still important. If we now, post-CMB, consider ourselves entitled to high-precision models and predictions for the physical state of the early universe and corresponding traces and relics, what methodological desiderata do we use to derive such predictions? What supports our extrapolating to the states of matter many orders of magnitude more extreme than anything we encounter in a laboratory? To answer these and similar questions, we need to shed light on the emergence and acceptance of the standard CMB interpretation and the rejection of the alternatives.

## 2. The microwave background phenomenology and its standard interpretation

The *standard interpretation of CMB* as *a remnant of the primordial fireball* was suggested by Dicke and his coworkers in their 1965 seminal paper (Dicke et al. 1965). While early predictions of Gamow and his students should not be discounted, the true history of the *physics* behind the CMB begins with Doroshkevich and Novikov (1964) and Dicke et al. (1965); in short, this was a set of ideas whose "time had come" and, hence, was taken seriously by both Soviet and American researchers.

The key idea behind the standard hot Big Bang interpretation is that in the early universe, baryonic matter (e.g., matter composed of protons and neutrons) and radiation were in equilibrium, similar to the equilibrium in human-made gas-discharge lamps. After the first light nuclei were formed (about 200s after the Big Bang) the universe was a hot, dense plasma of photons, electrons, and light ions, mainly protons and $^4$He nuclei. The plasma was opaque to electromagnetic radiation due to strong Thomson scattering by free electrons, as the mean free path each photon could travel before encountering an electron was very short. A high ratio of photons to baryons ($\sim 10^9$)– in those early stages (and today) makes this Big Bang model *hot*.

Eventually, the universe expanded and cooled to the point where the formation of neutral hydrogen was energetically favoured, and the fraction of free electrons and protons compared to neutral hydrogen rapidly decreased in a process known as *recombination*.[3] A similar process took place for helium, albeit somewhat earlier, with an intermediate state of singly-ionized helium $He^+$. This caused a strong scattering of photons by free electrons (Thomson scattering) to be substituted by interactions of photons with neutral atoms, processes weaker by many orders of magnitude. This substitution is equivalent to switching our gas-discharge lamp off. Soon afterwards, photons decoupled from (baryonic) matter, and their mean free path became comparable to Hubble length, or roughly the size of our cosmological domain, and the universe became transparent to light. This occurred about 400,000 years after the Big Bang, at redshift $z \approx 1100$. Photons from this epoch – or incoming from the *surface of last scattering* – travelled and cooled down freely, without interacting with matter, until some were stopped by a horn-shaped antenna in Holmdel, New Jersey, in 1964.

---

[2] The experimentalists often do not fare significantly better; to mention one illustrative example, by overlooking an alternative theoretical model, CERN physicists failed to detect a new kind of mesons their apparatus was readily producing (Pais 1986, 97). See Perovic (2011) for similar examples.

Apart from the background nature of those cosmological photons (they predate most other sources in the universe), it was clear that as matter in the universe is isotropically distributed on large scales, this primordial radiation must be distributed the same way. Since it originated in the opaque state of plasma, it should not be modulated and should behave as a perfect blackbody.[4] Finally, due to the expansion of the universe, the temperature of this blackbody should have decreased from the original thousands of kelvins at recombination to a few or a few dozen kelvin *The plausibility of the hot Big Bang model and the interpretation of the CMB, along with the plausibility of alternative models, were to be decided based on precisely how much that temperature decreased.*

The prediction of such a primordial relic was made, in fact, in the context of the relativistic hot Big Bang developed by George Gamow and his team long before the discovery by Penzias and Wilson, but it was largely ignored. Gamow and his students Ralph Alpher and Robert Hermann first understood around 1946 that a very hot initial state must have been opaque and that the subsequent recombination would cause photon decoupling and the emergence of a cosmological photon reservoir with adiabatically decreasing temperature (Gamow 1949). Alpher and Hermann predicted what later gained fame as the "temperature of the universe" as about 5 K (Alpher and Herman 1948a, b; 1949). Subsequently, both they and their teacher Gamow revised the value. Their predictions for the temperature were, however, inherently uncertain since they used the equilibrium Saha approximation (Gamow used Jeans stability criterion, inapplicable in such a simple form in the presence of dark matter) which had a poorly understood value of the Hubble constant, so their own varying values ranged between 5K and 50K.[5] While widely known, this prediction was not taken seriously before the "great controversy" forced astronomers to search for observational tests of world-models.

---

[3] As is the case with many labels in astronomy, this is a misnomer: as electrons and ions met in the cosmological context for the very first time, it should properly be called *combination*. While it would be ludicrous to suggest changing the terminology after so many decades, this curious little misnomer points to a much larger epistemological issue in history of science: the elapsed time until misnomers and circularities appearing in the heat of controversy become regularized and ironed out in the "normal" scientific practice. In a sense, this applies to the "Big Bang" as well; Hoyle's metaphor gradually lost both its pejorative sense and the sense of misnomer.

[4] Of course, this is valid up to the order of minuscule anisotropies reflecting the beginnings of structure formation; while this is of paramount importance for *present-day cosmology*, the small-scale anisotropies were unobservable for decades between the 1965 discovery and the advent of *COBE* in 1990s and, consequently, played no role in the overarching interpretation.

It did not help the case of Gamow and his collaborators that the non-equilibrium calculations required in the cosmological case require researchers to numerically solve differential equations on a computer, a method which was still uncommon in the 1940s and 1950s. Those calculations were finally performed independently by Dicke and Peebles at Princeton and Doroshkevich and Novikov in Moscow between 1960 and 1964 (Doroshkevich and Novikov 1964; Dicke et al. 1965). The initially obtained values were too high, at about 40K, and were reduced ten-fold with Penzias' and Wilson's momentous discovery.

In general, the fact that a sufficiently sophisticated theoretical framework already existed contributed to a fairly quick and wide acceptance of what became orthodox interpretation.[6] Over the past several decades, this interpretation has gradually become an integral part of the standard cosmological paradigm. In fact, *abandoning this interpretation is equivalent to the rejection of the entire paradigm*. Our aim here is not to give a comprehensive account of the history of CMB research. Rather, we have assigned ourselves the much less ambitious task of attempting to understand the emergence of the consensus and the failure of alternatives; for fuller historical surveys, we direct the reader to some of the references.[7]

---

[5] In comparison to the present-day understanding, Gamow and others picture the transition between the radiation-dominated and matter-dominated universe far too late at far too low an equilibrium temperature. This is a consequence of assuming a too high baryon density of the universe; we now know that baryons comprise only about 5% of the total mass-energy budget.

[6] Another issue, in an ironic twist of history, was raised by none other than Sir Fred Hoyle, the proponent of steady-state cosmology; he said the CMB had indirectly been observed as far back as 1941 (!) by Andrew McKellar, a Canadian astronomer working at the Dominion Astrophysical Observatory in British Columbia. His observations of the rotational excitation of the cyanogen (CN) molecules toward the star ζ Ophiuchus clearly showed something exciting those rotational transitions. Assuming that it was a contact with the thermal bath, McKellar (1941) estimated the "temperature of deep space" to be about 2.3K. The uncanny coincidence of this estimate with the modern value for the CMB temperature has been confirmed by contemporary studies (e.g., Roth, Meyer, and Hawkins 1993) that find it is indeed the CMB which excites the lowest rotational levels of interstellar cyanogen.

[7] The historical entry points are discussed by North (1994) and Partridge (1995), and a detailed theoretical introduction is given by Stebbins (1997). The history of measurements of the CMB temperature up to the early

As we see it, there were four stages in the establishment of the orthodox interpretation of the CMB. First, the theoretical model of the hot Big Bang was developed starting in the 1940s, with various consequences and parameters that were, at least in principle, observationally testable. Second, precise calculations of empirical consequences of the model, including the background radiation, were performed in the 1960s. Third, the discovery by Penzias and Wilson – as interpreted by Dicke et al. (1965) – was quickly and increasingly seen as a successful test of the hot Bing Bang model. *A moderate convergence of agreement* on the model and the interpretation of the discovery within it began in this stage. The fourth stage started when the *COBE* satellite observations were overwhelmingly interpreted as identifying the true origin of the CMB. This quickly resulted in *a wide convergence of agreement* on the hot Bing Bang model and the interpretation of the CMB within it. In a sense, as different aspects of the CMB were discovered, they gradually constrained alternatives to different extents, with the postulation of alternatives becoming particularly challenging after the *COBE* discoveries.

Yet between the initial moderate convergence and the wide convergence that followed, a number of alternatives were developed, defended, criticized, and repositioned, to a great extent with respect to the content developed in the four crucial stages of the emerging orthodoxy. Thus, to explicate and assess the alternatives, we first need to identify the crucial aspects of the orthodoxy itself.

Great interest in the properties of the CMB has led to innumerable theoretical and observational studies over the last decade, following in the wake of the extraordinarily successful *COBE* mission. One of the most impressive recent results is the measurement of the CMB temperature at the epoch corresponding to z = 2.34 using properties of the molecular hydrogen in a damped Ly-alpha absorption system in the spectrum of background QSO (Srianand, Petitjean and Ledoux 2000). The obtained result, although characterized by a large error margin, corroborates the predictions of the standard CMB interpretation.

The following three observationally established properties of CMB are crucial for any explanatory attempt:

1. **Spectral shape:** CMB is a blackbody to very high precision levels (Mather et al. 1994). See **Figure 1**.

---

1980s is given in Chapter 12 of Narlikar's work (1983). General accounts of the status of the modern cosmological paradigm can be found in many advanced textbooks; the one most frequently used in our study is by Peebles (1993), but see also Weinberg (2008).

2. **Temperature:** CMB has, according to the complete *COBE* dataset, a temperature of 2.728 ± 0.004 K (95% confidence level; Fixsen et al. 1996) or 2.72548 ± 0.00057 K (*COBE* dataset with *WMAP* recalibration; Fixsen 2009) or 2.7260 ± 0.0013, according to the still improving *WMAP*+*Planck* dataset (Hinshaw et al. 2013).

3. **Isotropy:** microwave background is fairly uniform over the sky. Except for the celebrated dipole anisotropy, explicable, as we shall see, through the motion of the observer together with the Local group, anisotropies have been detected only recently, using *COBE*, at large angular scales (7° and larger) and at the extremely faint level of $\Delta T/T \sim 10^{-5}$ (Smoot et al. 1992; Hinshaw et al. 2003, 2013). See **Figures 2 and 3**.

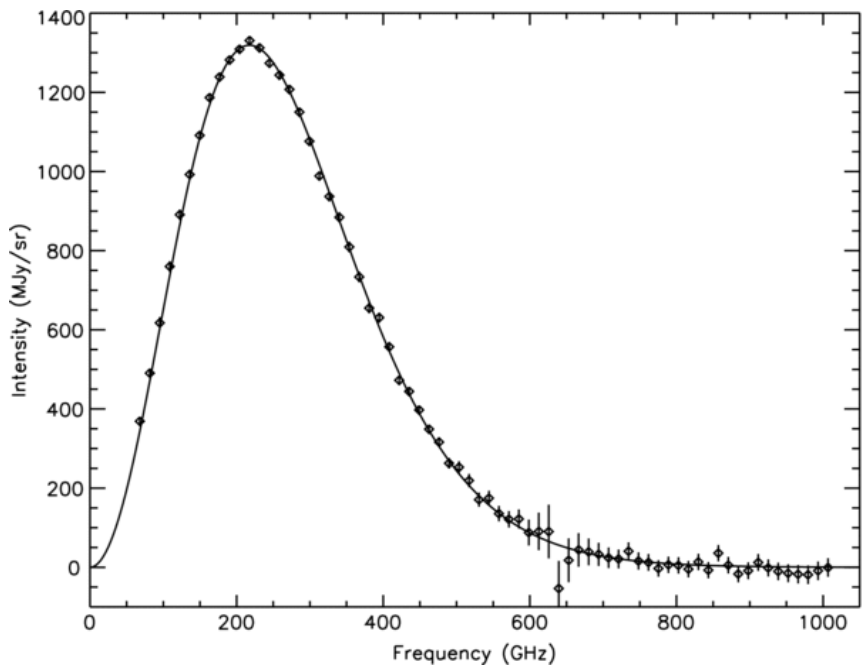

**Figure 1.** Blackbody spectrum of CMB as established by the FIRAS experiment on board *COBE* with *WMAP* recalibration (Fixsen 2009). (Courtesy of NASA.)

Obviously, these properties are not completely independent. Thus, it would make no sense to talk the global CMB temperature if it were not for its blackbody shape and unusual isotropy. In a sense, we can regard − following the standard paradigm − spectral shape and isotropy as the primary properties, and the temperature as, at least in principle, an adjustable parameter. Through one of the particularly "lucky" contingencies of history, Richard Tolman, in his influential 1934 book − many decades before the discovery of the CMB – proved the Hubble expansion of the universe would preserve the blackbody shape of any initially present blackbody radiation, with only the temperature decreasing linearly with the scaling factor. (As shown below, this physical fact makes the standard CMB interpretation "natural" but interferes with some of the attempts to interpret the radiation as a patchwork of sources thermalized at different epochs.)

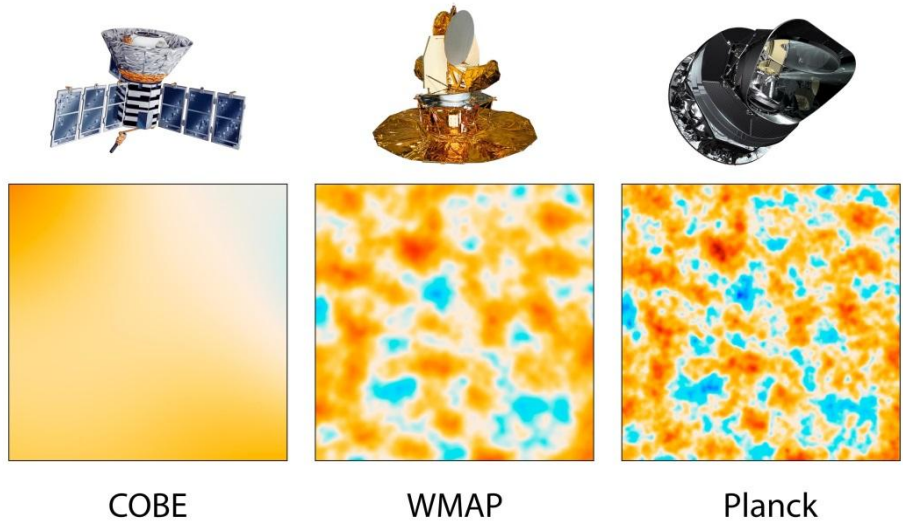

COBE          WMAP          Planck

**Figure 2.** The advancement of space-based CMB observatories. While *COBE* discovered intrinsic anisotropies in CMB (those which are not consequences of our motion), only with the advent of *WMAP* and *Planck* were we able to obtain insight into the *map* of the CMB, opening a whole new era in observational cosmology. (Courtesy of NASA/JPL-Caltech/ESA)

Probably the most significant and most frequently cited *consequence of the standard hot Big Bang interpretation of the CMB* is the limit the background temperature sets on the fraction of universal density which can be in the form of baryonic matter. The physical picture underlying this prediction is simple: the baryonic number is (at least approximately at the timescales comparable to the Hubble time, neglecting effects of the hypothetic proton decay and other very slow processes) a *conserved* quantity, and the vast majority of photons currently existing in the universe are CMB photons,[8] so the photon-to-baryon ratio today is essentially the same as it was at the time of decoupling, at redshift $z \approx 1100$. Therefore, fixing the photon density per co-moving volume, coupled with limitations on the baryon-to-photon ratio in the early universe (provided by the theory of primordial nucleosynthesis; Copi, Schramm, and Turner 1995; Schramm and Turner 1998), gives a unique handle on the total cosmological baryon density $\Omega_b$.

---

[8] In an early study, Shakeshaft and Webster (1968) show the energy density ratio of primordial to non-primordial radiation is about 400 to 1. This conclusion is independent of the interpretation of the CMB; for instance, although in the steady-state universe the total number of photons emitted by conventional sources such as stars within a sufficiently large co-moving volume diverges, as does the number of thermalized photons originating with a hypothetical early (Pop III) stellar population, providing both first metals and the CMB energy. In fact, Sir Fred Hoyle often uses this "coincidence" to argue for the Pop III origin of CMB (e.g., Hoyle 1994).

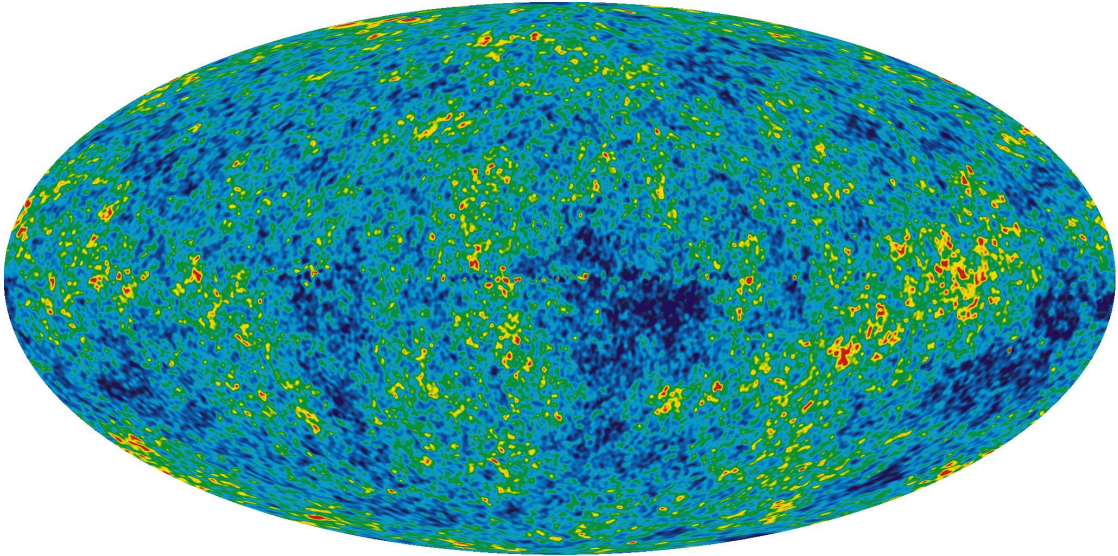

**Figure 3.** *WMAP* all sky survey of CMB anisotropies. The Internal Linear Combination Map minimizes the Galactic foreground contribution to the sky signal. It provides a low-contamination image of the CMB anisotropy, which translates into the angular-scale power spectrum of primordial inhomogeneities. It is, arguably, the major tool of contemporary cosmologists. (Courtesy of WMAP Science Team and NASA)

Now, turning to the taxonomy of the CMB unorthodoxies, we propose dividing them into two broad categories. The first category consists of those predicated on the *acceptance of the cosmological validity of field equations of general relativity*. These interpretations operate within the framework of Friedmann models, although this is not necessarily their explicit or even desired limitation. Prototypical examples are cold or tepid Big Bang models (e.g., Carr 1977; Carr and Rees 1977; Aguirre 1999). We will label interpretations belonging to this category as *moderate unorthodoxies*. The second category of *radical unorthodox* interpretations includes accounts of the CMB within *non-relativistic world models*, like the various steady-state cosmologies. Some brief historical discussions of CMB interpretation are scattered throughout the literature. Layzer and Hively (1973), before discussing their own unorthodox interpretation of the CMB data available at the time, explicitly divided known interpretations into two groups, approximately along the same lines as we do here. But no systematic and comparative treatment – especially not with the current *WMAP/Planck* data – has yet appeared. Ours represents the first such analysis.

Our division roughly corresponds to the division between *astrophysically* and *cosmologically motivated* interpretations. The details of this division will become clear in what follows, as we consider each interpretation in more detail. In fact, modern cosmology abounds with problems tackled from two different points of view: astrophysics and particle physics, and its newest child – the nascent discipline of quantum cosmology. This divide has recently acquired not only methodological traits but sociological ones as well.

## 3. Moderate unorthodoxies: CMB as a relic of Population III objects

The prime examples of *moderate unorthodoxies* are developed within the cosmological models of cold or tepid Big Bang. Such models are variations on the standard theme of the singular origin of the universe in relativistic Friedman models, but under different initial conditions, particularly a low or intermediate value of the photon-to-baryon (or entropy per baryon) ratio η. As we discuss below, these models have some conceptual advantages over the standard hot (= high value of η ∼ $10^9$) Big Bang. Also, non-standard models in the context of CMB interpretations are bouncing universes in which the Big Bang is not regarded as a singularity, but as a bounce from the previous, contracting epoch; while mostly ignored today as obsolete, bouncing models have enjoyed brief episodes of popularity, notably in the 1970s and 1980s.[9] All these non-standard cosmological models have varying degrees of difficulty accounting for the CMB origin. They have enjoyed a measure of support and have been regarded as viable contenders by at least some individual cosmologists for most of the last half century, especially immediately after the great cosmological controversy in the late 1960s when the failure of the theory of the steady-state universe left us with a single grand cosmological narrative. And even today, variations on the theme of non-standard Big Bang occasionally appear in the most authoritative astrophysical journals (Aguirre 1999, 2000; Li 2003; Yi-Fu, Easson, and Brandenberger 2012); these variations are not necessarily, of course, motivated by any CMB-related considerations.

---

[9] See Kragh (2013b) for early history of these models. Classical bouncing models ran into problems with both the entropy production in the epoch preceding the bounce and the lack of a persuasive physical mechanism for the bounce itself. Some similar models are now proposed in string and loop-quantum cosmologies based on particular approaches to quantum gravity.

There have been several attempts to build the model of Big Bang with low photon-to-baryon ratio; as noted, these are often known as cold or tepid Big Bangs (Zeldovich 1972; Barrow 1978; Layzer 1968, 1992; Aguirre 2000).[10] But why would anyone prefer a cold Big Bang? An initial advantage of such models is that they offer a significantly easier ground for the formation of the initial perturbations which would grow to become today's galaxies and the current large-scale structure. In addition, through the so-called Kibble mechanism, they *a fortiori* avoid several problems related to the phase transitions in the early universe, like the overproduction of monopoles or other topological defects (cf., Kolb and Turner 1990).[11]

It is not surprising, then, that until very recently, these models retained some degree of popularity and plausibility, with many attempting to assimilate the CMB results into them. For example, Aguirre (1999, 2000) maintains the non-primordial CMB formed by Population III at high redshifts ($z \geq 100$) and moderate temperatures could plausibly be thermalized to within the current observable limits if there is enough conveniently shaped and abundant dust. However, as there is no early thermal plasma in these models, there are no acoustic waves propagating through plasma, and, consequently, there can be no acoustic peaks in the angular spectrum of CMB anisotropies. Therefore, the discovery of acoustic peaks by *WMAP* (and some of the balloon CMB experiments, like *BOOMERANG*; Crill et al. 2003) refuted cold or tepid Big Bang theories.[12]

The fact that these theories are no longer viable contenders for an overarching cosmological paradigm actually makes our forensic analysis easier: we can separate those theoretical elements which address the issue of CMB origin and analyse them independently from the rest of the framework. In other words, we need not enter into discussions of broad advantages and disadvantages of cold Big Bang models, but can concentrate on what is arguably the most direct feature of such models, the necessity to distinguish a physical mechanism separating the CMB origin from the Big Bang. Simply put, if we accept the premise of a cold or tepid Big Bang, we need to find a source of more than 99% of photons currently existing in the universe, one that is *distinct* from the Big Bang itself. Such sources are hypothetical Population III stars.

---

[10] Sometimes these models are inadequately labelled Big Bang models with low entropy (e.g., Penrose 1979).

[11] We explicate scientific motivation for some of the interpretations, but a separate study focusing on the motivations of the cosmologists to introduce their interpretations is needed.

[12] For example, Komatsu et al. (2011). We note in passing that this is one of the clearest instances of the Popperian falsification in recent scientific practice: a set of unique and causally necessary predictions has been found to be in gross conflict (formally, many hundreds of standard deviations) with clear-cut experimental results.

The expression "Population III" is used here in a more liberal sense than in most technical astrophysical discussions. The conventional view is that this population represents the first generation of stars, with zero metal content (e.g., Carr 1994).[13] Yet any objects – not necessarily Main Sequence stars – forming in early epochs and becoming the first to reach (quasi)stability may be said to comprise Population III in a wider sense. Thus, pregalactic stars and black holes, even magnetoturbulent pregalactic gas clouds (Layzer 1968), also belong to this category.[14]

Some alternative cosmologies strive to explain the CMB without a dense, hot phase near the beginning. Completely cold Big Bang models can be distinguished from the "tepid" version using the nucleosynthetic criterion. As Carr (1981b) writes:

> In proposing that the 3K background is generated by pregalactic stars or their remnants, we are not necessarily assuming that there are no photons before the stars form. This would be unrealistic since many processes in the early Universe would inevitably produce some radiation (e.g. the dissipation of initial density fluctuations or primordial anisotropy). Furthermore, a Universe which was "cold" at the neutron-proton freeze-out time (~ 1s) would be unlikely to produce the observed 25 per cent helium abundance through cosmological nucleosynthesis (although the helium might, in principle, be synthesized in pregalactic stars).

The cold Big Bang models have not, in general, been met with such profound hostility and merciless criticism by the mainstream cosmology as the steady-state and later Hoyle-Narlikar models. Therefore, a comparable amount of research has not been invested in discovering all the implications of and possible weaknesses in such models. The view of these models as "toy models," the main purpose of which is to highlight the advantages of the standard model, has been reinforced by the fact that some of the main and most authoritative protagonists of such unorthodoxies soon switched to the standard paradigm (Zeldovich, Rees,

---

[13] These would be stars with primordial chemical composition (only hydrogen and helium, with negligible amounts of $^7$Li), initiating all subsequent chemical evolution of baryonic matter.

[14] The physical meaning of the notion is sometimes blurred. We need to keep in mind that the size of the first objects to undergo gravitational collapse is highly disputed issue. Researchers usually calculate the Jeans mass for each epoch in the early universe. There is some controversy over the entry parameters, as well as the true role played by the Jeans mass in the formation of compact objects (e.g., Gnedin 2000; Schneider et al. 2003). All these characterizations of Pop III objects are therefore provisional, pertaining to the local goals of the relevant study; they are not part of the core of relevant knowledge. And Pop III objects are logically necessary within the standard cosmological picture as well; after all, *some* metal content had to be formed first in such objects. Their properties and prospects for detection are vigorously debated and remain an area of active research.

Carr). Yet we do not know whether such a view is entirely justified, given that these "toy models" have never been considered and developed as genuine models of our universe.

The fact that these non-cosmological explanations of the CMB origin have sometimes required special physical conditions which are to be evaluated separately complicates things even further. In practice, the issues of developing a cosmological model on the one hand, and considerations of special physical conditions on the other, have traditionally been conflated. It is thus easy to fall into a trap – to regard the presence of specific sorts of dust in galaxies or intergalactic media as *an inherently cosmological* issue, postulated by an overarching cosmological concept as analogous to matter creation in the classical steady-state theory. However, they are clearly phenomena of a local astrophysical nature for which plausible local astrophysical explanations are required.

Two additional related are worth considering. First, there is *a possibility that a non-negligible fraction of the observed CMB energy density originated in the primordial fireball and that alternative origins should be considered only for the rest*. This partition is present in several models put forward in the literature; we discuss these in the following sections.

Second, *it is often possible to accommodate alternative sources of the CMB photons within a predominantly primordial origin picture*. In such cases, since there is no *a priori* reason to expect a perfect fit of two physically distinct radiation components, the issue of distortions and deviations from the perfect primordial blackbody comes into play. There are two basic types of distortions: comptonization (or Sunyaev-Zeldovich) distortions measured by parameter $y$ and chemical potential (or Bose-Einstein) distortions, usually denoted by $\mu$.[15] The latter is usually thought to be zero, in accordance with the limits set by *COBE* (or is unobservable, like the $\mu = 3 \times 10^{-10}$ inferred by Smoot et al. 1992). The comptonization $y$-parameter, however, is acknowledged as an important measure within the standard CMB interpretation and is the key to understanding subsequent, astrophysical distortions, such as the Sunyaev-Zeldovich effect in rich clusters of galaxies or the effects of dark energy on small-scale CMB anisotropies. The reference value for the present discussion is again the *COBE* value of $|y| < 2.5 \times 10^{-5}$ (Mather et al. 1994); this will be also important for later discussions. While this huge topic is beyond the scope of the present study, the historical process leading to consensus about these deviations is of considerable interest and has not been studied.

---

[15] The "minimal" comptonization parameter $y$ describes a redistribution of photons in the reference blackbody spectrum, usually via Compton scattering, but this could be generalized to any form of energy release which does not result in the higher temperature blackbody, but in spectral distortion. It is another observable property of the CMB which obtains a natural interpretation (Sunyaev-Zeldovich) within the standard paradigm, but could be problematic for some of the unorthodoxies, as, for example, the model of Gnedin and Ostriker.

A final issue is the appearance of non-desired non-primordial origin of part of the CMB photons in schemes designed to account for completely different observations or theoretical presuppositions. Such is the case, for example, in Gnedin and Ostriker's (1992) high baryonic density universe. This model is motivated, in part, by a desire to solve the well-known problem of dark matter in galaxies. The authors take great pains to show that the observed smoothness of the CMB is consistent with the primordial origin in the framework of their model in the same manner as in the standard low baryonic density models; the only outcome of very massive Population III objects they postulate is a Sunyaev-Zeldovich distortion parameter $y \sim 10^{-4}$. While this can be considered falsified in the aftermath of *WMAP* and *Planck*, it is important to understand that this was clarified only in the last 10 years.

### 3.1. CMB as discrete sources: Rowan-Robinson

In the early days after the discovery of CMB radiation, it was impossible to observe its "fine structure," i.e., to achieve high angular resolution in observations. The natural idea in such cases (repeated in a similar form in the controversy still surrounding the X-ray background; see, for instance, the review of Fabian and Barcons (1992)) is that instead of truly diffuse emission, more or less uniform over the entire sky, we are dealing with many discrete sources, too numerous to be separated by our observational techniques.

It is hard to sufficiently emphasize how poor – in comparison to today's post-*WMAP*, post-*Planck* era – the observational database of early CMB studies was in the late 1960s and throughout the 1970s. Even the dipole anisotropy (anisotropy due to the motion of the observer, i.e. the Solar system, relative to the background) was controversial until the famous experiment of George Smoot and co-workers with a differential radiometer mounted on a U-2 spy plane (Smoot, Gorenstein, and Muller 1977; Gorenstein and Smoot 1981).

An early idea about the CMB being due to the superposition of sources was suggested by Gold and Pacini (1968), more as a plausibility argument than as a serious astrophysical hypothesis. The interest in such models peaked around the end of the 1960s (Wolfe and Burbidge 1969; Alexanian 1970; Smith and Partridge 1970; Setti 1970). It later declined, and no new models of this kind were proposed after 1974. This is not surprising. In contrast to the models directly linking CMB origin with the hypothetical Population III of pregalactic or early galactic objects with subsequent thermalization, the discrete source models depended in a straightforward and testable manner on the existence of a rather narrowly defined class of astrophysical microwave sources, and the explosive development of radio astronomical techniques in the early 1970s completely ruled this out.

In a short paper written in 1974 , British astrophysicist Michael Rowan-Robinson proposes a simple discrete source model for CMB origin which nicely illustrates some of the most basic points of this unorthodox tradition (Rowan-Robinson 1974). Well-written and clear, the article paints a simple picture using the (still scarce) observational data on the CMB at the time. However, subsequent observations, particularly our new knowledge of the short-wavelength part of the CMB spectrum, soon made this model obsolete, making it an unusually straightforward example of falsificationism in scientific practice. Rowan-Robinson ascribes the energy of CMB photons to *known sources*, Seyfert and other types of active galaxies, which were already well-known and intensely investigated by the early 1970s.

Rowan-Robinson significantly contributed to the development of what is today called the Standard Model of active galactic nuclei (e.g., Rowan-Robinson 1977). Their spectrum is approximated by a simple broken power law. This is a particular advantage of this theory in the Popperian sense, as further investigations of the spectral energy distribution of these sources might well disprove the theory. Furthermore, the theory does not claim the sources should superpose in such way as to completely mimic the blackbody spectrum. As shown in Figure 1, the predictions of Rowan-Robinson's model begin to differ from the standard ones at log ν > 11.3, with the offset sharply increasing toward larger frequencies.

An early argument against many superposed sources as the origin for CMB photons (irrespective of the issues of thermalization), due to Hazard and Salpeter (1969), states that the anisotropies to be expected on purely statistical grounds for randomly distributed sources should, of course, be in the order of $1/\sqrt{N}$ , where *N* is the number of sources per unit solid angle (or the beam size). This value, for the number of galaxies shown in existing surveys, should be about 0.6% and should be contrasted with the relative amplitude of CMB anisotropies, *then unknown, but constrained* to be less than 0.1% (excluding the dipole anisotropy). Background discrete sources must be significantly more numerous than the known galaxies to produce such remarkable uniformity of the CMB. The introduction of populations of hitherto unknown sources whose properties are solely determined through the postulated origination of CMB photons represents the introduction of *ad hoc* explanatory entities. Even though this may undermine their plausibility, it is fair to say that such entities appeared earlier in the history of science (e.g. phlogiston or aether) and again more recently (e.g., dark energy and dark matter).

Interestingly, the Rowan-Robinson model uses Milne's (Special Relativistic) cosmological model finding, because "if the microwave background is due to sources, the evidence that General Relativity applies on a cosmological scale is not compelling" (Rowan-Robinson 1974, p. 46P). This is a significant piece of reasoning, especially because it comes from a proponent of the very cosmological outlook he is questioning. In fact, two years earlier, Rowan-Robinson entitled an editorial comment in *Nature* "Steady state obituary?," referring to the most prominent non-relativistic cosmological framework. However, it seems that he − among, no doubt, many other cosmologists in the early 1970s − felt the orthodox explanation of the CMB origin is the most important argument, the only crucial one for the hot Big Bang paradigm. Rowan-Robinson, thus, is a good example of a scientist who, while accepting the paradigm, continued to investigate. Although he subsequently wrote on related issues, such as distortions of the CMB spectrum by dust (Rowan-Robinson, Negroponte and Silk 1979) or microwave emission by various classes of extragalactic objects (Ade, Rowan-Robinson and Clegg 1976), these were indirectly or not at all relevant to the issue of the origin of CMB photons.

Naturally, Rowan-Robinson's particular model of discrete sources was disproved soon after its formulation. It says something about the prevailing atmosphere in interested circles that it did not merit a separate paper containing a detailed refutation. The very shape of the spectrum at the high-frequency limit is enough to disprove this model, as became clear with the measurements of Smoot et al. (1987) and their claims that the blackbody shape is satisfactory in the wavelength range of 0.1-50 cm.

## 3.2. Thermalization by grains: Layzer and Hively

Layzer and Hively give a detailed discussion of the CMB origin in their 1973 proposal of a non-standard cold Big Bang model. David Layzer started investigating various cosmological and cosmogonical ideas in the mid-1950s. In particular, he was interested in the problem of galaxy and large-scale structure formation; he argued a cold Big Bang is a more hospitable environment for rise and development of density fluctuations of the required magnitude (Layzer 1968).

In the 1973 model, Layzer and Hively postulate that early Population III stars are the ultimate source of energy of CMB photons − and the above comment by Carr fully applies here. Hydrogen fusion in stars provides sufficient energy only if shining stars comprised a much bigger part of the universe in early epochs than they do now. In other words, in order for stellar-powered CMB to work, dark matter today should not only be baryonic, but also should consist mainly of stellar remnants (presumably old neutron stars and non-accreting black holes) of those early populations. The theory of primordial nucleosynthesis was not well-developed in 1973, however, and Layzer and Hively could not give specific predictions of maximum baryon abundance consistent with the observed amount of light nuclei; thus, their gross overestimation of $\Omega_b$ was understandable.

The redshift at which the bulk of the CMB energy was released is not tightly constrained in the Layzer and Hively model: it allows for the contribution of the gravitational collapse of primordial gas clouds, which could occur only at $z \geq 10$, while the massive Population III stars could release bulk of their energy somewhat earlier, by $z \leq 50$.[16] The interval $50 \geq z \geq 10$ is the best anyone could do in cold Big Bang models without having further physical insight into the nature of the Population III sources. Once again, what Hoyle and others repeatedly touted as an advantage of unorthodox alternatives – outsourcing the explanatory work to better-known stellar astrophysics – turned out to be an explanatory burden.

We now arrive at the key issue of thermalization of Population III starlight (or infrared luminosity from collapsing gas clouds). According to the observations available at the time, CMB retains its perfect blackbody shape up to wavelengths of about 20cm. If that is due to opacity of dust, then the universe must be opaque at wavelengths below about $20(1+z)^{-1}$ cm, where $z$ is the poorly-determined redshift of the energy origination. This requires huge amounts of intergalactic dust grains, with huge amounts of metals existing to the present day, leading to all sorts of difficulties, including the extinction of light from distant quasars and galaxies. If grains are roughly spherical with the standard values for dielectric constants, then the fraction of total cosmological mass locked in grains must be enormous: $\Omega_{dust} > 0.06$. This is about an order of magnitude larger than the amount of matter in all luminous stars.

In addition to the spherical grains, Layzer and Hively (1973) discuss what they call a somewhat bizarre case: grains in the shape of hollow spheres (i.e., spherical shells). They offer a plausible qualitative model of the formation of the strange grains: the formation of a carbon and silicate shell around an ice core, followed by sublimation of ice through the porous shell. Layzer and Hively mention, but only in passing (and credit the comment to a private communication from Purcell), the possibility that the grains may be elongated. This idea of this particular form of thermalizing agents gained popularity in the work of Hoyle and Wickramasinghe.[17] As these later researchers point out, the strangely shaped grains could reduce the total mass of dust and metals required. Yet not only is the reduced mass still too large for realistic assessments, but it takes exotic methods to produce such strange grain shapes.

---

[16] We should remember here and elsewhere that the conversion between redshifts and ages (or distances) occurs differently in different cosmological models.

The model of Layzer and Hively *requires* that $\Omega_b \approx \Omega_{tot} \approx 1$. Although it does not explicitly exclude any form of non-baryonic matter, it obviates the necessity for it. Although the total matter density can, in principle, be significantly higher than the critical density, thus enabling a significant quantity of metals to be condensed in thermalizing grains, it is not necessary (and would be problematic for other reasons). However, values for $\Omega_{tot}$ much smaller than unity like the one usually obtained in observational surveys are strongly excluded, since the small fraction of baryons leads to a still smaller fraction of metals; hence, the density of the thermalizing grains is too small. This is an important prediction, especially because a strong revival of interest in large quantities of dark matter in both the Western world and in the then Soviet Union came somewhat later than the Layzer and Hively study (Ostriker, Peebles and Yahil 1974; Einasto, Kaasik and Saar 1974).

David Layzer returned to the topic of cold Big Bang in the last decade of the 20[th] century in a short paper on large-scale structure formation (Layzer 1992). In the paper, he re-emphasizes the advantages of a low photon-to-baryon ratio from the point of view of the growth of density perturbations. Clearly, he has not abandoned all hope of finding a method of efficient thermalization to account for non-primordial CMB origin.

## 3.3. Primordial chaos: Rees (I) and Eichler

In an early paper, Sir Martin Rees (1972) proposes an additional source of energy for the CMB: the dissipation of primordial chaotic fluctuations. This is not a "true" thermalization model, so we consider it separately from the later model of Rees (1978).

We knew much less about the primordial power spectrum in 1970s, so various scenarios were concocted. Perhaps the most popular category fell under the umbrella term "chaotic cosmology," although the attribute "chaotic" was not, obviously, linked to issues of predictability and computability as it is now, especially in nonlinear dynamics and celestial mechanics. In one important sense, the meaning of "chaos" at the time was antithetical to the one in widespread use now. Primordial chaos served contemporary cosmologists mostly as an *ansatz* whose role was to remove the dependence on initial conditions and to show that various kinds of initial conditions lead essentially to the same universe at later epochs (thus, for instance, Misner 1968). Clearly, that is very different from the notion that chaos is the property of systems *highly sensitive* to initial conditions and that a small difference in the initial conditions leads to exponential divergences in the subsequent evolution (e.g., Strogatz 2001). Rees uses primordial chaos in operational terms, as a placeholder for initial inhomogeneities and anisotropies whose smoothing released the required CMB energy. He points out that in the standard picture, the redshifts of decoupling of matter from radiation and recombination have to be very close; he regards this as an unnecessary explanatory burden. Rees shows how primordial irregularities of different mass scales could provide continuous input of heat at the expense of gravitational potential and kinetic energy, until the size of the horizon grew sufficiently to encompass the region of the universe we consider to be "sufficiently" homogeneous. He is frank about the fact that his proposal is rough and speculative: "We leave the physical details of the dissipative processes as an open question" (Rees 1978, p. 1670).

---

[17] It is not clear why Layzer and Hively do not discuss this case (arguably less "bizarre" than the spherical shells) in some detail.

Irrespective of the source, thermalization is easier at early epochs, *ceteris paribus*, as optical depths are huge, even within the context of open baryonic universes used by Rees ( $\Omega = \Omega_b < 1$ ). The degree of "clumpiness" is an important theoretical parameter fixing the smallest redshift at which the thermalization may occur. Parenthetically, this is the only alternative model for CMB origin which postulates the redshift of origin *higher* than in the standard pictures: Rees estimates that the bulk of energy was released and thermalized by $z \geq 10^4$ .

This idea is tackled by an Israeli astrophysicist, David Eichler (then at the University of Chicago), in a short paper discussing the role of entropy fluctuations in the early universe prior to formation of the earliest visible structures (Eichler 1977). Eichler adds several interesting elements to Rees' picture, notably that the dissipation of turbulences through shock waves is the most important mechanism for the input of heat. Dissipation proceeds until the low mass of $M \sim 10^6$ $M\odot$ is reached. He connects this to the formation of first globular clusters, the oldest existing relics of structure formation. The hypothesis of Rees and Eichler could, therefore, be regarded as a "secondary unorthodoxy" following the primary one of chaotic cosmologies. Subsequent events leading to the demise of the chaotic cosmology program (see, for instance, the account in Peebles 1993), obviated the need for further research along these lines, although some important explanatory concerns appear again, disguised as modern inflationary theories.

We should emphasize that although from both historical and epistemological points of view, cosmological inflation is not a logically necessary part of the standard hot Big Bang cosmology, it has become a welcome and extremely fruitful expansion of the standard model in the last 35 years. In particular, *WMAP* and *Planck* have shown that particular properties of the CMB anisotropies map are well-accounted for by the inflationary framework (e.g., Komatsu et al. 2011). Theoretical quibbles notwithstanding (e.g., Ijjas, Steinhardt, and Loeb 2014; Linde 2014), the inflationary paradigm has brought us much closer to understanding initial conditions of the structure formation process and promises tighter integration with other branches of fundamental physics, including conformal quantum field theory and string theory.

## 3.4. Thermalization and the return of large numbers: Rees (II)

In 1978, Sir Martin Rees returned to the scene of models for a non-primordial CMB origin with a modernized version of Layzer and Hively's hypothesis (Rees 1978). The motivation for this particular Population III model is, however, different than for his initial model. The main problem Rees tackles is the origin of the above-discussed photon-to-baryon ratio. He finds the situation in which η could be derived from astrophysical processes and constants preferable to the one in which it has an obscure, albeit cosmological, origin. It seems obvious that such motivation is deeply intertwined with the controversy over fine-tuning physical constants and cosmological parameters for habitability (e.g., Barrow and Tipler 1986; Barnes 2012). In Rees' words: "Here a possible non-primordial origin of the microwave background is outlined that seems less contrived than other such schemes" (p. 35).

This model hinges on the assumption that early pregalactic sources radiate at the Eddington luminosity, a critical value where radiation pressure balances gravity. For accreting black holes, it occurs naturally, but Population III *stars* have to be supermassive for the approximation to hold. But if it holds, a general argument links photon-to-baryon ratio η with the fourth root of the ubiquitous Dirac's "large number" $10^{40}$ multiplied by a string of model-dependent factors of the order unity. Since it is related to the properties of the Eddington luminosity, then, η is not *ad hoc* in this model but is tightly constrained by "usual" astrophysics.

Another important feature of Rees' model is the involvement of two other processes, re-radiation in molecular bands and free-free absorption by the ionized intergalactic medium. While the first process requires too many molecules to be efficient, free-free absorption and scattering will create a large optical depth at a longer wavelength, even for small concentrations of free electrons $n_e$ maintained by photoionization (and presumably, though Rees does not dwell on it, shock ionization caused by formation, mass-loss, and explosions of at least some of the pregalactic Population III stars). Rees admits that "the value of $n_e$ is too model-dependent to permit any firm estimate of how important this process is" (1978, p. 36). He gives the redshift of the bulk of CMB formation as $z\sim100$ or somewhat larger, much earlier than we now observe.

Overall, Rees' model received much more publicity than Layzer and Hively' model, appearing even in a number of popular or semi-popular works (e.g., Barrow and Tipler 1986) and garnering a reasonably high number of citations over time.[18] However, as with other moderate unorthodoxies requiring many luminous Population III sources, its validity hinges on the large baryonic cosmological density and strange initial mass function of those early stars. In a very similar model proposed by Hayakawa (1984), star formation must be strongly suppressed *below 200 Solar masses* (!) to avoid the explosive ejection of metals and the degree of chemical enrichment prohibited by observations. Both Hayakawa and Rees admit that the presence of non-baryonic dark matter (and/or dark energy) inflicts a strong blow on any such endeavour.

### 3.5. Late thermalization of normal starlight: Rana

Indian astrophysicist (and student of Narlikar) Narayan Chandra Rana (1981) gives a model with by far the latest thermalization. In several respects, this is the most radical of the moderate unorthodoxies discussed in this section. It does away with *both primordial and Population III origin* of CMB photons, proposing instead the thermalization of "normal" starlight at redshifts in the range $10 \geq z \geq 5$. Since the starlight energy density, as noted above, is much smaller at present, Rana's model also requires strong starburst activity at said epochs.

---

[18] NASA Astrophysics Data System (http://www.adsabs.harvard.edu/) give it as 109 citations (as of June 6, 2016), corresponding to moderately successful 2.9 citations per year.

Rana's model has several appealing features. First, it does not have the isotropy problem characterizing other models for CMB origin, including the Standard Model. As is commonly known, this problem is one of the most important motivations of modern inflationary models (see Linde 2008). More specifically, in Rana's model, the angular size of the horizon at, say, $z \sim 8$ (with realistic values of the deceleration parameter $q_0$) is large and may easily comprise the entire sky. This means the isotropy of matter distribution – the cosmological principle of Eddington and Milne – within our visual horizon implies the isotropy of the total light contribution. In addition, this model produces the necessary helium in stars instead of in the primordial nucleosynthesis. Rana develops a fairly general formalism for the thermalization of the cosmological population of sources and the temperature evolution of radiation. The formalism does not put any extravagant constraints on the nature of the Population III sources.

In this sense, Rana's model is the least assuming of the moderate unorthodoxies. This means its failure highlights all fatal flows of the entire project of decoupling the CMB from the Big Bang and the primordial epoch. Not only are thermalizers postulated *ad hoc*, as Rana honestly admits, but specific chemical composition (graphite) and shape (long needles or whiskers) are required for the thermalization to succeed. The influence of Hoyle, Narlikar, and Wickramasinghe is obvious in this respect (see Section 4), although, to Rana's credit, his mathematical model is fairly general. It tends to overestimate the magnitude of small-angular scale fluctuations in the CMB temperature for about an order of a magnitude (this is understandable as the model was constructed a decade before *COBE*). It also favours an open, low-density universe with the best-fit deceleration parameter of $q_0 = 0.12$; this was falsified, however, after 1998 and the emergence of the new standard cosmological model.

## 4. The CMB in radical unorthodox models (without Big Bang)

The classical steady-state theory of Bondi, Gold and Hoyle was very much alive at the time of the discovery of the CMB. The theory already had several distinct problems, mostly with the radio source counts, as well as recently discovered high-redshift objects, QSOs, but these obstacles did not seem insurmountable. An excellent account in a monograph by Kragh (1996) shows how the steady-state paradigm had managed to overcome seemingly serious observational refutations, like the (spurious) Stebbins-Whitford "effect" prior to 1965. It would not be abandoned without a struggle.

In this section, we discuss various explanations of the CMB that follow in the steady-state tradition. This tradition continues to offer the most persistent challenge to cosmological orthodoxy, quite understandably, given its main protagonists: Sir Fred Hoyle, Jayant Narlikar, Geoffrey Burbidge, and a few others. The battle has been fought on two fronts: the authors note weaknesses and problems with the standard paradigm (Arp et al. 1991); they propose a revised (or quasi-) steady state theory (Hoyle, Burbidge and Narlikar 1993, 1994; Narlikar et al. 2003). We find two different attempts to account for the CMB photons: first, thermalization of distant discrete sources by some form of cosmic dust grains and, second, divergent scattering at the "domain boundaries" by particles, mostly electrons, with variable mass. The former is characteristic of the attempts to account for the CMB in both classical and revised steady-state theories, and in some more moderate unorthodoxies like the cold Big Bang models discussed in Section 3. The latter is relevant only for a specific form of Hoyle-Narlikar conformally invariant cosmological model (see Section 4.1.).

Other radical unorthodoxies discussed in the section are the plasma cosmologies of Alfven and Lerner and the closed stationary models of Ellis, Maartens and Nel (1978) and Phillips (1994a, b).[19] We do not deal with those unorthodoxies lacking anything particularly original or interesting to say about the CMB phenomenon, including various fractal/chronometric cosmologies, Dirac's large-number hypothesis or tired-light models (but see Sorrell 2008). At best, these rehash the explanatory mechanisms developed by steady-state (or cold Big Bang!) proponents, notably thermalization of background sources on dust grains. In particular, the tired-light models – in which photons lose energy either by interacting with the intergalactic medium or by travelling through a vacuum, i.e. "on their own," leading to the claim that the universe is not really expanding – have been refuted and are even regarded as hallmarks of pseudoscience.[20]

## 4.1. Changing masses: Hoyle

In response to the downfall of the classical steady state theory, during the late 1960s and 1970s Sir Fred Hoyle and his student Jayant Vishnu Narlikar developed an array of new theories based on the general idea of large-scale stationarity. Although the new theories were simply variations of the classical steady state concept, in the views of the authors, they represented a "radical departure" from the previous steady state concept (Hoyle and Narlikar 1966). They centred on extensions of Hoyle's field theory version of the classical steady state cosmology to produce *a conformally invariant cosmology*. It is indicative that the other two fathers of the classical steady state concept, Sir Hermann Bondi and Thomas Gold, publicly disavowed any association with later Hoyle-Narlikar versions.

---

[19] A detailed theoretical classification of radical cosmological unorthodoxies up to early 1980s can be found in the review of Ellis (1984).

[20] One of the most cogent formulations is given by LaViolette (1986). For classical refutations, see Alpher (1962), Zeldovich (1964) and Wright (1987); empirical arguments up to the mid-1980s are summarized in a comprehensive review by Sandage (1988); for a recent negative test, see Foley et al. (2005).

The explanation of the CMB origin in the 1972 version of these theories (Hoyle & Narlikar 1972a, b) is provocative and intriguing. It is discussed in detail in a beautifully written paper by Hoyle (1975).[21] Basically, by extending the ideas of Wheeler-Feynman classical direct-particle interaction model from electromagnetism to gravitation, Hoyle and Narlikar create an unorthodox theory of gravity. The key idea is that, while dimensionless quantities are all fixed, we should be able to express any dimensional quantity using particle masses and these dimensionless quantities – and, hence, translate any dynamic aspects of the evolution of any system into changes in the masses of particles over spatiotemporal coordinates. It turns out that the application of Hoyle-Narlikar's gravity to cosmology requires variable masses of particles as functions of their spatial coordinates.

If we accept such variation of masses, it is clear that under fairly general conditions, at some point in spacetime, all the masses will become negative. This is not very disturbing, as it is only important to have all the masses in a single causally connected region of the same sign.[22] Thus, we may imagine the universe consisting of two halves, with opposite signs of the masses and opposite signs of the mass field contributing to the action described above. By further generalizing this picture, we may get a whole net of aggregates (one of the first multiple-universe schemes in cosmology!) instead of two, as shown in Fig. 1 in Hoyle's 1975 paper.

Obviously, the interface between the two halves of the Hoyle-Narlikar universe is the *zero-mass surface*. Strange physics taking place in this region because of vanishing particle masses may create a simulation of the physical condition near the Big Bang in the Friedmann cosmologies, from the point of view of a distant observer. In particular, the scattering amplitudes tend to infinity, as they are inversely proportional to the masses of scattering particles. This pertains to electrons, which scatter any photons present extremely efficiently. A very large (formally divergent) amount of scattering is bound to produce the exact blackbody spectrum, indistinguishable from what has been observed. In this respect, the zero-mass surface in Hoyle-Narlikar's theory corresponds to the surface of the last scattering in the standard hot Big Bang cosmology or to a homogeneous distribution of thermalizing grains in models with thermalization of distant sources (like the Layzer and Hively model).[23] Any amount of matter, even a minuscule one, near the zero-point surface will act as a perfect thermalizer – and if we stick to the cosmological principle of homogeneity and isotropy within each domain (as Hoyle and Narlikar do), there is no reason to expect any dearth of matter near the boundary. Thus, an interesting – or bizarre – consequence of Hoyle-Narlikar's theory is that all primary anisotropies should be *exactly zero*. In stark contrast to both the standard model and all other alternative hypotheses for the CMB origin, the spectrum will always remain a featureless blackbody. This is the reason for Narlikar and Rana's (1980, 1983) claim that the Hoyle-Narlikar theory offers a better fit to the CMB – in pre-*COBE* times – than the standard cosmological model. Of course, the advent of *COBE* and other experiments revealing complex and extremely informative structure in small-scale anisotropies falsified this radical prediction (Wright et al. 1994).

---

[21] See also Narlikar and Rana (1983).
[22] One of the appealing feature of any such scheme is that it automatically answers a philosophically interesting question which bothered Einstein, namely why – in contrast to electromagnetism – all charges to which gravitation is coupled (= masses) have the same sign.

Unfortunately, Hoyle's (1975) paper has received little attention.[24] Obviously, the Big Bang orthodoxy became much stronger after the CMB discovery. Other arguments on the nature of gravitational interactions in the conformally invariant cosmology led to the theory and the cosmological superstructure it suggests being considered highly implausible. Despite the relativism of any aesthetic choices in science, however, it is our impression that the elegance and subtlety of the Hoyle-Narlikar theory is unmatched by other CMB explanations, whether orthodox or unorthodox.

## 4.2. A revised steady-state

In the early 1990s, the last and most comprehensive instance of the classical steady state theory was formulated under the name of quasi-steady state (henceforth QSS; also sometimes called revised steady state) theory in a series of papers by Hoyle, Burbidge, Narlikar and occasionally Arp, Wickramasinghe, Sachs, and collaborators. Without going into the wealth of technical details (presented *in extenso* in Hoyle and Burbidge 1992; Hoyle et al. 1993, 1994; Narlikar et al. 2003), we should mention that as in Hoyle's version of the classical steady state model, the negative energy of the creation field (C-field) transforms into matter with positive energy. However, the creation is not uniform in spacetime but occurs in discrete creation events, the so-called "mini-bangs." In each individual "mini-bang," about $10^{16}$ Solar masses (a characteristic mass of superclusters of galaxies) are created in the form of particles with Planck mass ($M_{Pl} \sim 10^{-5}$ g). The distribution of creation events creates the characteristic cellular structure seen in large galaxy surveys of recent decades.

---

[23] This is not *exactly* true, as the effective surface of last scattering will be provided by whatever surface *close* to the zero-mass surface is capable of producing a spectrum deviating from the perfect blackbody by our instrumental uncertainties. This is not just splitting hairs. As the effective thermalizing agent, the zero-mass surface and its vicinity are stronger than any other such agent conceived by any other alternative hypothesis discussed here. Therefore, Hoyle's hypothesis cannot be rejected on the basis of insufficient thermalization, in contrast to Rowan-Robinson's or Rana's models.

[24] Among 34 citations reported on the NASA Astrophysics Data System by May 2016 – a remarkably small number for a paper published 41 years ago, by such a famous author and on such a hot topic – one is a self-citation, one is a book chapter devoted to Hoyle's achievements, six (17.6%) are by a single author (Canadian cosmologist Paul S. Wesson, with collaborators) interested in the variability of fundamental physical constants, and three deal with the issue of CMB origin; none of these was published after 1980.

The scaling factor in QSS cosmology has an exponential (as in the classical steady state) and an oscillating component, typically of the following form:

$$R(t) = e^{\frac{t}{t_c}} \left[ 1 + \varepsilon \cos\left( \frac{2\pi\tau}{Q} \right) \right].$$

(This could be derived from C-field dynamics; see, e.g., Hoyle et al. 1993.) Here, $t_c$ and $Q$ are two timescales, one for the conventional Hubble expansion, and the other for the temporal amplitude of mini-bangs, while $\tau$ is a function of cosmic time $t$ which deviates from $t$ only near minimal values of $R(t)$ – near the local "minibangs." Those epochs are characterized by the creation of new matter, ultimately in gaseous form. Parameter $\varepsilon$ has absolute value less than unity, so the scale factor never actually reaches zero.

As far as the CMB origin is concerned, the answer of Hoyle et al. is essentially the same as earlier thermalization propositions, although, of course, there is no universal Population III in QSS. A somewhat novel idea is that thermalization is carried out in two phases, first by carbon whiskers converting starlight into infrared radiation and next by iron whiskers ("needles") producing the observed microwave background. The only component of the integrated starlight which cannot be entirely thermalized is the starlight originating in the last generation of supercluster formation, i.e., the last minibang. As Narlikar et al. point out, "These will stand out as inhomogeneities on the overall uniform background" (2003, p. 2), like the small angular scale anisotropies discovered by *COBE*.

Simply stated, this cannot work, and Wright et al. (1994) and Wright (2003) indicate many fatal problems with this account. Notably, the metallic whiskers required for thermalization would have caused a huge optical depth of the order of 100 (!) on millimetre wavelengths towards sources located at a redshift of about 2. The very fact that we readily observe some such sources among IRAS superluminous galaxies, for example, argues against the QSS scheme. In addition, the implied power spectrum of perturbations is incompatible with both *COBE* and *WMAP* data, predicting the spectral index of $P(k) \propto k^n$ to be $n = 3$, while the *COBE* value is $n = 1.2 \pm 0.3$, and the *WMAP* nine-year dataset value is $n = 0.972 \pm 0.013$. No mechanism suggested thus far could remove those discrepancies, while retaining a remote chance for QSS.

Two 21[st] century studies of interest to this alternative view (or family of views, as QSS has been modified several times since its proposal by Hoyle, Narlikar, and Burbidge) are by Li (2003) and Fahr and Zönnchen (2009). Li (2003) returns to the topic of the convenient "needle" shape of intergalactic dust necessary as a thermalization agent, not only for QSS, but for a host of other alternative models. Improvements to the theory of the extinction of electromagnetic waves are substantial, although still not sufficient to determine whether a convenient form of grains could be found. Even so, it is instructive to briefly compare this work with the early discussions of dust thermalization cited above (e.g., Layzer and Hively 1973; Wickramasinghe et al. 1975; Rowan-Robinson et al. 1979; Rana 1979, 1980; Carr 1981a,b; Bond et al. 1991). Dielectric dust grains seem to be conclusively rejected now, whatever shape we consider; only speculative conducting ("metallic") elongated grains ("needles") remain even remotely viable. Their tenuous viability is undermined by the fact that the "widely adopted Rayleigh approximation is not applicable to conducting needles capable of supplying high far-IR and microwave opacities" (Li 2003, p. 598). No one has proposed an adequate model (*Ibid.*):

> Due to the lack of an accurate solution to the absorption properties of slender needles, we model them either in terms of infinite cylinders of according to the antenna theory. It is found that the available intergalactic iron dust, if modeled as infinite cylinders, is not sufficient to produce the large optical depth at long wavelengths required by the observed isotropy and Planckian nature of the CMB... the applicability of the antenna theory to exceedingly thin needles of nanometer/micrometer thickness needs to be justified.

It seems we are back where we started; by the beginning of the new millennium, the search for thermalizing dust grains had become what Lakatos (1978) calls degenerative research.

This is seen in the paper by Fahr and Zönnchen (2009). These authors critically review major results of the CMB astrophysics and propose retreating to what is essentially the QSS framework. They recycle an old argument of Rees (1978) on the dimension analysis of the baryon-to-photon ratio, this time without proper emphasis on the fact that fusion/accretion must proceed at the Eddington limit to provide the necessary CMB energy. Ironically but fittingly for this rear-guard action, they find that even the entirely hypothetical metallic whiskers are not enough for thermalization to satisfy modern observational constraints – so they invoke the ancient tired-light yarn (photons losing energy as they interact with the matter while travelling through the static universe) as part of the explanation.

All in all, QSS theory comes across as similar to the cold Big Bang cosmology in its explanations of the CMB origin and must face the same objections as the models of Layzer and Hively, Rees (II), and Rana.[25] Despite the valiant efforts of Hoyle, Narlikar, Wickramasinghe and others, even *ad hoc* postulated thermalizing agents are incapable of saving the theory.

## 4.3. The CMB in the plasma cosmology

A particular type of unorthodox cosmological model, *plasma cosmology*, was proposed by Swedish physicist Oscar Klein and developed and defended with great vigour by his compatriot, Nobel-prize laureate Hannes Alfvén (e.g., Alfvén and Mendis 1977; Alfvén 1979). It reached a wide audience through its popular exposition in a book by Eric Lerner (1991), also a plasma physicist. Simply stated, plasma cosmology argues for symmetry between matter and antimatter, with rather slow annihilation, which could, in principle, provide the energy contained in the CMB (and much else).

In several papers, mostly published outside the "core" astrophysics *magisteria*, and in his book, Lerner tries to formulate an alternative account for the CMB properties (Lerner 1988, 1995). This is mostly a rehashed version of the story of *helium + thermalizing dust* production in early supermassive stars, as discussed in Section 3. There is, however, one novelty in Lerner's account: he claims the intergalactic medium is a strong absorber of radio waves. This absorption is supposed to occur in narrow filaments, with tiny holes scattered about randomly so that distant compact radio sources like QSOs and radio-galaxies can be seen through the holes: "Thus, if our hypothesis is valid, the microwave photons we see were last scattered a few million years ago, no 15 billion years ago" (Lerner 1988, p. 464). Lerner (wisely) refrains from discussing how much baryonic matter must be locked in those intergalactic filaments, nor does he say the total real population of QSOs and similar sources must be large to account for the size of the observed subpopulation (if we see a dozen people in a room while peeking through a randomly positioned narrow keyhole, it is reasonable to conclude that the room contains very many people).

---

[25] See also Kragh (2012).

The irony is that, in the end, Lerner cannot sustain the explanation based on plasma cosmology alone and takes refuge in tired-light ideas, eerily similar to Fahr and Zönnchen (2009)![26] This, more than anything else, demonstrates how low the stock of Big Bang opponents has fallen since *COBE*.

## 4.4. Closed steady-state models and the CMB: Ellis et al. and Phillips

*Closed steady-state models* were developed after 1965 by theoretical cosmologists with no serious desire to challenge the Big Bang orthodoxy. Instead, they wanted to achieve a better understanding of the underlying postulates of the Friedmann models and their relationship with the empirical data.

For instance, the closed static model of Ellis and his collaborators (Ellis 1978; Ellis et al. 1978) challenges the observationally unverifiable postulate of the large-scale homogeneity of the universe. In this model, which aims at studying global assumptions like the cosmological principle, the universe is isotropic only around a specific point in space; we are located in close proximity to this point, because of the anthropic selection effect. Following the weak anthropic principle, it is expected that we are located in regions possessing the necessary properties for the creation and evolution of complex biological systems, near the "centre" of the universe (it is most natural to use this term for our pole of the manifold, by analogy with a 3-sphere). A singularity surrounded by hot matter is located opposite the centre, simulating the initial singularity in the Friedmann models. However, in this static model, the singularity is *co-present* with everything that exists; it does not precede it. Obviously, this makes the model more appealing from an epistemological point of view, although the laws of nature break down at singularities. In Ellis et al.'s model, the singularity is not always inaccessible in the past; it could be, in principle, investigated using the methods and apparatus of modern science.[27] This co-present singularity is easiest to intuitively understand as "an enclosure" or "a mantle" surrounding the universe. Its major purpose is to be a recycling facility in the global cosmological ecology, as the static nature of the universe makes the recycling of high-entropy matter necessary to explain the obvious and "anthropic" observation that the universe is not in a heat death state. In the framework of Ellis et al.'s model, this is achieved by the streaming of high-entropy matter (mainly in the form of heavy elements synthesized in stellar nucleosynthesis) toward the singularity, where it is dissociated and returned to the universe in the form of low-entropy matter, presumably hydrogen atoms or plasma of baryons and leptons. Other than this streaming, which does not change the net mass distribution, there is no systematic motion: all observed redshift is entirely of gravitational origin.

---

[26] See http://www.astro.ucla.edu/~wright/lerner_errors.html, last accessed June 8, 2016.

Ironically, although this model is highly non-standard and was rejected from the start because of its inability to properly account for the redshift-magnitude relation known since Hubble, its explanation of the CMB origin is the *standard* one rehashed. The CMB photons originate in close spatiotemporal proximity to the global singularity, and that proximity is – in Ellis et al.'s model, as well as in the standard cosmology – hot and optically thick. The only difference is that we can speak of a *continuous production* of the CMB in the static model of Ellis et al., while in the standard picture, and all other alternatives considered above, the CMB was produced in a definite past epoch of the universal cosmic time. Although an interesting phenomenon in itself, the physical mechanism of matter recycling near the singularity has not been investigated in detail.

As emphasized by the authors of this unorthodoxy, such a model can already be considered observationally disproved. The original paper by Ellis et al. (1978) shows this model cannot account properly for the (*m, z*) curve, i.e., the relationship between the apparent magnitude and redshift of cosmologically distributed sources of radiation. Since this is one of the most basic facts of observational cosmology there is no doubt that this hypothesis is falsified – and as Ellis repeatedly suggests, its point was to better elaborate the notions of singularity and cosmic entropy *within the standard picture*. For our purposes, we only need to note that the reasons for its rejection are not directly linked to the CMB or its properties.

---

[27] This could be read as a late echo of ideas promoted in the time of the classical steady state theory, especially Sir Hermann Bondi's endorsement of an extreme Popperian view of falsifiability, which would discard any considerations of unobservable cosmological phenomena, such as those beyond our particle horizon or before the universe became transparent at recombination; see Bondi (1955, 1992) and comments in Kragh (1996).

The same applies to a similar model by Phillips (1994a, b). This model is somewhat more complicated as it includes two different kinds of matter; there are also two singular points, this time called the northern and southern pole. The Milky Way galaxy is located in close proximity to the northern pole of the universe (for the same anthropic reasons as in Ellis et al.'s model). In contradistinction to the Ellis et al. model, this one proposes the systematic motion of galaxies from the northern to the southern pole. The motion, however, is laminar and appears stationary, so that the universe, in general, always offers the same picture to a typical observer. Because of this large-scale motion, the observed redshift is partially of gravitational and partially of Doppler origin. It is harder to disprove Phillips' model than Ellis et al.'s model (1978), as the gravitational and Doppler redshifts are delicately entangled in the former. The cleanest test could be the measurement of a peculiar motion of distant sources with respect to the universal reference frame as defined by the microwave background radiation. This measurement is possible in rich galaxy clusters by means of the Sunyaev-Zeldovich effect (Sunyaev and Zeldovich 1980). The prediction of the Phillips' model is that more distant clusters will tend to have significantly larger peculiar motions than nearby ones, counter to the standard theory of structure formation with cold dark matter. Recent measurements indicate this is not the case, however, and there is no meaningful way to save the theory (Phillips, private communication to one of the authors).

However, as far as the CMB origin is concerned, these models do not go far astray from the main idea of the standard model; i.e. they associate the CMB origin with the physical state of matter close to the global singularity (only the nature of this singularity is different in the standard lore). Hot plasma near the singularity plays the role of an almost perfect blackbody, and its radiation is redshifted in one way or another to form the observable CMB. While in the standard model, the CMB is created once, and while this creation lies deep in our past (as measured by universal cosmic time), in the model of Ellis et al. the source of the CMB is co-existent with us.

The very existence of these models, not to mention their internal theoretical consistency, demonstrates the falsity of the prejudice often repeated by uncritical supporters of the standard paradigm that only Big Bang theory offers a *natural origin* of the CMB photons. In addition, it shows that our basic dynamical theory in cosmology, namely the theory of General Relativity, allows many more solutions if the cosmological principle of homogeneity and isotropy (which is partly a metaphysical conjecture) is abandoned. The role of inhomogeneities in the observed CMB is less significant in these models than in standard cosmology. Their stationary nature puts the issue of structure formation on an entirely different physical footing.

## 5. The formation of the orthodoxy and the alternatives: an epistemological framework

We have seen several interesting trends in the dissenting tradition of CMB origin. In the incrementally emerging mainstream cosmology, i.e., the hot Bing Bang model and a corresponding CMB interpretation (see Section 1), in the first few decades after the discovery, no individual "heresy" came close to attracting wide attention. The most seriously debated unorthodox solution to the CMB origin puzzle is likely Rees' model (1978). It is useful, as it contains most of the alternative ideas and mechanisms in a nutshell. A substantial shift in dealing with alternative ideas occurred from the mid-1970s to the mid-1990s. The shift is best explained by the incoming *COBE* results, but the less obvious and more gradual acceptance of large quantities of non-baryonic dark matter and the relegation of baryonic matter to a secondary role are other factors.

When they did discuss the alternatives, the defenders of the emerging orthodoxy almost uniformly preferred to deal with one "heretical" idea at a time and to avoid quoting similar ideas. These are clear signs of a careful theoretical debate of a significant issue. There was never any sign of summary dismissal or witch hunts, however, and the debate never spilled into the news media with all the public acrimony that might follow.[28]

There was also an absence of consensus among challengers, unlike the emerging orthodoxy. We can at least partially explain this by a widespread impression that the Big Bang scenario itself is inviolable; it may be slightly modified but not radically rejected so there is no need for a unified alternative. This does not apply to the indefatigable Hoyle or, at least partially, to the school of younger Indian astrophysicists inspired by him (Narlikar, Wickramasinghe and Rana). Insofar as some degree of opposition consensus emerged, it not only happened in a haphazard fashion, but it was also fixed on the most speculative and *ad hoc* aspect of alternative theories, graphite and metallic whiskers as the thermalizing agents. Nor can we observationally sample interstellar or intergalactic dust for grains of a particular kind (and will be unable to do so for quite a few centuries to come). The consensus looks very much like tailoring reality to suit the theory. This may be true of the orthodox account as well but, to make the situation worse, there was no theoretical consensus that the proposed grains would do the explanatory task *even if they really existed*, as noted in a careful study by Li (2003).

---

[28] We could argue, in fact, that there was more bad blood between some of the main supporters of the standard cosmology; one example is. the infamous split between main *COBE* investigators and subsequent Nobel Prize winners John Mather and George Smoot over the latter spilling the information to news media (Mather and Boslough 1997).

And the dissention about the origin of the CMB has never truly become a controversy. Scientific controversies are typically stages in the development of science characterized by unusually prominent political or ethical disagreements entangled with the theoretical and factual disputes. They are typically expected to be resolved by careful appeal to facts and sharpened theoretical principles (Engelhardt and Caplan 1987). The discussions of the CMB have never contained any unusual social and methodological elements, in sharp contrast to the great cosmological controversy of 1950s and early 1960s between the classical steady state theory and the Friedmann models which would become the standard cosmology. With the exception of theories suggested by Hoyle, Narlikar, and Wickramasinghe – which stemmed from much wider unorthodox cosmological schemes of steady-state or "quasi" steady-state ideas – all other unorthodox CMB theories are firmly entrenched in the same milieu as the modern mainstream cosmology outlined at the beginning of the paper. Plasma cosmology has never become a serious contender, and closed steady state models were from the beginning conceived as thought experiments rather than attempts at explaining physical reality. Almost all of the proponents of the hypotheses presented in Section 3 fully agreed on a set of major premises: a) conservation laws are universally valid, and the underlying dynamical theory is Einstein's general relativity; b) the universe is, on average, homogeneous and isotropic; c) the universe started in some form of a Big Bang.[29] This may not be surprising: despite a wide range of diverse views of the relevant phenomena they covered, the sources of most unorthodox CMB theories were observational facts (e.g., the deficit of baryonic matter compared to the nucleosynthetic constraints, or the existence of hierarchical cosmological structure) completely uncontroversial and fully admitted by mainstream scientists as difficulties, or at least phenomena in need of explanation.

The plan of discussing only one heretical idea at a time changed to wholesale criticism as new observational data solidified support for the standard paradigm and weakened rival interpretations. One particularly salient example is the effect the arrival of *COBE* data had on proposals on the origin of the CMB. A study by Wright et al. (1994) discusses (and rejects) a wide range of unorthodox ideas and models. Some are moderate unorthodoxies, like the high-baryon universe of Gnedin and Ostriker (1992), while others are clearly unorthodox proposals, notably the early thermalization model of Layzer and Hively discussed earlier. Practically all unorthodox CMB explanations were obviated by the set of *COBE* data, and even minor contributions of non-primordial origin became severely limited. By the time of *WMAP* and *Planck*, nobody bothered to refute the remnants of troubled unorthodoxies.[30]

---

[29] As mentioned in Section 3.1, Rowan-Robinson (1974) is an exception; he used Milne's special relativity cosmology rather than Friedmann models as the background. However, there was no particular motivation for this. He explicitly notes that the Milne model "is a good approximation to a low density $\Lambda = 0$ Universe with matter" (p. 46P), i.e., an open Friedmann model.

In a sense, the *COBE* experiments brought an end to the "controversy that was not," with a convergence on the hot Bing Bang cosmological model and the fireball interpretation of the CMB. In retrospect, we can see the diversity and varying ramifications of unorthodox theories. The lack of such a controversy matching the great cosmological controversy on the cosmological models before the *COBE* results came in masks the fact that the earlier debate was substantial and careful, and the formation of the orthodoxy was not an immediate done deal. This is not surprising, as other convincing empirical reasons that could trigger overwhelming consent simply did not exist. In fact, the debate was driven by theoretical insights and preferences that, on their own, could hardly deal a devastating blow to competing alternatives. Rather, careful theoretical consideration of each of the models prepared the way for the consensus established when the *COBE* evidence concurred with the interpretation regarded as theoretically best motivated. In other words, the alternative models played a decisive role in the formation of the orthodoxy that only in retrospect may seem to be an independently developed approach.

Perhaps this theoretical building-up of consensus was partially due to the fact that the wiggle room for alternative interpretations is much wider in cosmology, as it is essentially observational science, than in experimental physics, as it provides much more direct evidence to opposing sides in debates. The underdetermination of theoretical accounts by evidence is bound to be much more pronounced and longer lasting in cosmology. The CMB was a milestone discovery but it would be misleading to expect it played a similar role to that, for instance, to the effect of the evidence of the existence of an elementary particle delivered by a collider on competing theoretical approaches. And it would be misleading to predicate an historical account on such a view.

In general, failing to understand the subtleties of the history of the establishment of orthodoxy runs the risk of eliciting widespread prejudice that there are only few insignificant opinions dissenting from the standard paradigm. In our case, even the bibliography of relevant work speaks volumes (literally!) on the unsoundness of this prejudice. Among the scientists connected with various non-standard hypotheses, are some of the most authoritative figures of 20th century astrophysics, including Sir Martin Rees, David Layzer, Geoffrey Burbidge, Jeremy Ostriker, and Sir Fred Hoyle.

---

[30] Again, with the exception of Ned Wright, whose web site (http://www.astro.ucla.edu/~wright/errors.html, last accessed June 9, 2016) contains important – up to 2010 – criticisms of the non-mainstream approaches.

The number of studies devoted to the standard interpretation is overwhelming. This is not necessarily a good indication of the number of alternative accounts, however, as most studies *explicitly assume* the standard interpretation is valid, *including those that reconsider various aspects of it*. Thus, we find much serious reconsideration of the key aspects of the orthodoxy – e.g., the relation between the CMB anisotropies on the one hand, and the quantity and the kind of structure in the standard λCDM model on the other [31] – sheltering under the umbrella of the paradigm's general acceptance. Similarly, serious considerations of the possibility that General Relativity may not apply cosmologically if the CMB is due to discrete sources appear in the work of key proponents of the orthodoxy.

The statement of unqualified acceptance is often a rhetorical device used to prevent hasty conclusions on the status of the orthodoxy by either professionals or the wider public. And this is the aspect of knowledge production in cosmology that sociological studies of the discipline should take into account. The situation may be a specific and not necessarily desirable form of "paradigm defence" in the Kuhnian sense used by mainstream cosmology. Given all this, it is understandable but perhaps regrettable that the number of textbooks in which alternative interpretations are even mentioned is insignificant; counterexamples are usually found in textbooks written by the "mavericks" themselves (e.g., there is an interesting and open-minded account in Narlikar 1983).

## 6. What of alternatives?

The story of the CMB origin offers insights into the nature of the progress of modern science – its good and bad points alike. The role of the empirical but unexpected discovery of the CMB as unravelling the deepest mysteries of the origin of the universe was immediately and widely recognized by almost the entire cosmological community, including most researchers with unorthodox views. In general, it helped persuade a large portion of the wider scientific community that cosmology is a serious, mature and firmly founded scientific discipline.[32] The cutting of the Gordian knot of the great cosmological controversy opened up new vistas in cosmology. The level of sophistication of modern work in cosmology, like the N-body simulations of galaxy formation, investigations of the power spectrum of density fluctuations or the correlation and autocorrelation functions of various collapsed structures, the search for primordial particle relics, or theories of the QSO absorption line systems, to mention only a few, would be simply impossible without the discovery of the CMB and the effort invested in its standardized interpretation.

---

[31] The model contains a cosmological constant (λ) associated with cold dark matter (CDM), assuming the General Relativity is correct at large scales.

Overall, dissent has served the lofty principles and ideals of scientific enterprise.[33] Alternative explanations of the CMB origin have, in most cases, been falsified, leading to a new problem-situation permitting the emergence of new views, such as inflation or quantum cosmology, since the 1980s. A dramatic transition from the original problem-situation set up by the steady-state challenge to the relatively poorly defined relativistic orthodoxy in cosmology, to a completely new level of high-precision cosmology and studies of the very early universe was occasioned not just by a single momentous, epoch-making empirical discovery, but also by a blizzard of theoretical activity surrounding it, including many iterations of conjectures and refutations, some of which we have noted here. This happened in four stages (see Section 1), so that incrementally, over the course of three decades, the alternatives' manoeuvring space was reduced.[34] Even so, the alternative explanations have an ongoing role to play.

We can take three methodological lessons from this history. First, most alternative models accounting for the CMB have never been fully developed and are not even close to the level of detail of the hot Bing Bang model. As we have suggested, this may have to do with the lack of consensus among the unorthodoxies – they just did not have enough common ground to pull the resources together. In addition, some alternatives were toy-model reactions to the model that was becoming dominant. Yet the level to which a researcher can improve an alternative model overall should not be underestimated. For example, Reese's model II was a thorough refurbishing of Layzer and Hively's appealing yet deficient model. Also, the account of the CMB within the steady state model and its better fit until the *COBE* data arrived demonstrates how far a comprehensive and elegant but unorthodox model can go in accounting for the key physical facts. And even an incorrect model can be helpful in identifying the weaknesses of unorthodoxy, as was the case with the isotropy of the CMB in early versions of the standard model.

---

[32] This is sometimes disputed, but a full analysis is beyond the present scope; cf. Disney (2000); López Corredoira (2014).

[33] There are many dangerous preconceptions in popular controversies of science and pseudoscience (such as those related to global climate change or universal vaccination), which could and should be easily dispersed by investigations of "science at work". Interestingly, the defenders of science often use misplaced and misguided arguments which are easily demolished by detailed analysis of the history of philosophy of specific case studies, such as we do here. For example, at least one influential web encyclopaedia explicitly devoted to "[a]nalyzing and refuting pseudoscience and the anti-science movement" regards "alternative cosmology" as belonging to the same pseudoscientific category as "alternative medicine" (http://rationalwiki.org/wiki/Alternative_cosmology, last accessed June 11, 2016). Apart from the similar sound, there is no parallel between the two in either an epistemological or an ethical sense.

[34] In this manner, the whole case study may be understood as a refinement of falsificationist views of scientific knowledge. (Popper 1972, 1992) The view has been taken seriously by astronomers and cosmologists (Kragh 2013a).

Second, it is possible to develop alternatives employing a piecemeal rather than a wholesale approach. One of the lessons of our analysis of the history is that a general cosmological framework can be clearly distinguished in the explanations of the CMB origins, and the latter can be considered and developed independently. Thus, there are a number of routes we can take to rethink the details of physical significance and only subsequently turn to a more general framework.

Third, we should bear in mind that there may be less apparent or alternative broad theoretical presuppositions lurking in the background of the orthodoxy, and these may motivate and make valuable certain alternative explanations of key physical phenomena, such as the CMB. Devising models and explanations of key physical phenomena to support them rests on the intricate interplay of theoretical presuppositions and selected observations. A particularly instructive case is the requirement for large quantities of dark matter in more recent cosmology; this fits well with the earlier postulation of thermalizing grains to explain the CMB as non-primordial in Layzer and Hivey's model. Similarly, with new evidence of particular facts, e.g. those pertaining to baryonic matter, alternatives that seemed unappealing suddenly become plausible. For all these reasons, it is instructive to nurture a scientific community that actively develops alternatives to the orthodoxy and allows bold conjectures and fringe models.

Another question to consider is what sort of an edge, if any, the mainstream interpretation of the CMB and the inflationary approach in general has over alternatives, given the widely accepted criteria of what constitutes scientific evidence. Astroparticle physics can provide some evidence of particle properties, as for instance, cosmological constraints on the number and masses of neutrinos (Steigmann and Strittmatter 1971). But the cosmological evidence is, on the whole, very different from the evidence provided in, say, experiments in solid state physics, and the crucial aspects of theoretical models are much more directly tested in the case of particle physics tested in particle colliders. To give an example, we can never test the primordial fireball hypothesis directly, as we can myriad other physical phenomena in a laboratory. To clarify the nature of the evidence upon which cosmology is based, researchers sometimes compare it to the sort of constructive evidence found in archaeology or palaeontology. We may well ask to what extent this sort of analogy is adequate. We may even doubt that cosmology meets these standards of evidential support because it is so indirectly related to the core of theoretical models and leaves them open to underdetermination to such an extent that it is questionable whether we can label the evidence supportive.

The reference to palaeontology is apt, at least in terms of the theory-evidence relationship, although cosmological evidence has some important advantages, such as much earlier attention to the so-called selection effects that can be detrimental to constructive evidence. Yet we do not think the viability of a scientific field will necessarily be questioned simply because it does not meet the stringent standards of evidence set by experimental physics (and probably only in some areas of it). But in terms of the standards of evidence, the distance between experimental physics and cosmology in general, including the CMB case and its interpretations, is much greater than that between various cosmological approaches (including astroparticle physics). The latter are all confined to a common standard of evidence which is substantially different than the standard applied in most particle physics. Now, theoretical accounts based on such kinds of evidence are generally prone to more or less reticent underdetermination, even if the rivals are treated as falsified with a great deal of certainty. Given this, it may be useful to avoid treating failed alternative interpretations as we would treat falsified alternatives in experimental physics – i.e. as theories to be straightforwardly discarded. Instead, it may be more advantageous to regard them as a resource of approaches that can potentially but realistically be revised and revived or as initial dips into a wider pool of possible alternatives. Even though the plausibility of the existing alternatives has diminished with the advance in detection techniques, the nature of the evidence leaves the field open to various revisionings of the existing alternatives or to the development of their various elements than in, for example, particle physics, where models and theories have been straightforwardly discarded in light of the experimental results.

Following this rationale, it may be useful to rank these unorthodox CMB explanations in terms of the plausibility and persuasiveness of their theoretical grasp and promise. Such a ranking may provide an overview of the alternatives and suggest a framework for understanding and judging them, perhaps even for developing some of their ideas.

Within the "radical" group, the arguably most interesting solution is given in the framework of the conformally invariant theory with variable masses (Hoyle 1975). Its solution for the puzzle of CMB origin – divergent scattering of photons when passing through the boundary of cosmological domains with different mass signs – is devilishly ingenious. It promises interesting physical insights into such crucial questions as the nature of mass or the epistemological possibility of equivalent descriptions of a single and only indirectly accessible physical event. There certainly are not many cases where the famous Italian saying of *se non è vero, è ben trovato*[35] is more appropriate. On the one hand, it is pity that the brilliance and authority of Sir Fred Hoyle have not motivated the development of his ideas.[36] On the other hand, the same cannot be said for the radical unorthodox static models of Ellis et al. and Phillips which are, in a final analysis, currently little more than curiosities, abundant in the history of any sufficiently rich and dynamic scientific field.

Among the "moderates," the idea of discrete sources creating the CMB has been easiest to discard, since it essentially depends on the status of experimental techniques. The latter have seen almost explosive development throughout the last several decades, especially since the discovery of the CMB firmly established cosmology as a respectable user of cutting-edge observational equipment. Therefore, we are dealing with a sort of bootstrap, so often encountered in young sciences: a bold hypothesis is given emotional support or preference by a majority of the scientific community *in spite of lack of direct support*, motivating tremendous observational and experimental efforts whose results solidify the support for the hypothesis, turning it into a paradigm.

Those unorthodoxies involving new physical elements (such as variants studied by Carr, dealing with Population III stars and early black holes properties) are more difficult to deal with, as shown by some of the examples in Section 3 (the rest can be found in the literature), but discussions of them remain fruitful to this day. This is true even if we consider them as somewhat beyond the cosmological mainstream. As we have shown here, the mainstream focusing is a highly complex and nonlinear process, with many studies "gathering dust on shelves" becoming incorporated into the mainstream. Stock examples include Yang-

---

[35] "If it is not true, it is well conceived."

[36] A tangential issue has been relevant to the attempts, mainly by Canadian astrophysicist Paul Wesson and his collaborators, to build a cosmological model of our universe embedded in classical spacetime of higher dimensionality (e.g., Wesson and Seahra 2001).

Mills theories, baryon number non-conservation in particle physics, or horizontal gene transfer in evolution/molecular biology.[37]

---

[37] Three anonymous referees have immensely contributed to the improvement of a previous version of this manuscript. The authors acknowledge support through grants ON176021 and ON179067 of the Ministry of Education, Science, and Technological Development.